\documentclass[aps,prb,twocolumn,reprint,superscriptaddress,floatfix]{revtex4-1} 
\usepackage[T1]{fontenc}
\usepackage[utf8]{inputenc}
\usepackage{mathptmx}
\usepackage{amsmath}
\usepackage{graphicx}   
\usepackage{siunitx}    
\usepackage{epstopdf}

\usepackage[colorlinks=true,linkcolor=blue,citecolor=blue,filecolor=blue,urlcolor=blue]{hyperref}

\newcommand{\nmr}{\textsc{nmr}}

\newcommand{\rf}{\textsc{rf}}
\newcommand{\nqr}{\textsc{nqr}}
\newcommand{\bnmr}{$\beta$-\textsc{nmr}}
\newcommand{\bnqr}{$\beta$-\textsc{nqr}}

\newcommand{\musr}{$\mu$\textsc{sr}}
\newcommand{\slr}{\textsc{slr}}
\newcommand{\lemusr}{\textsc{LE}-$\mu$\textsc{SR}}
\newcommand{\eli}{\textsuperscript{8}{Li}}
\newcommand{\elip}{\textsuperscript{8}{Li}\textsuperscript{+}}

\newcommand{\nli}{\textsuperscript{9}{Li}}
\newcommand{\nlip}{\textsuperscript{9}{Li}\textsuperscript{+}}

\newcommand{\lip}{{Li}\textsuperscript{+}}
\newcommand{\sto}{SrTiO$_{3}$}
\newcommand{\lao}{LaAlO$_{3}$}

\newcommand{\triumf}{\textsc{triumf}}
\newcommand{\isac}{\textsc{isac}}
\newcommand{\cmms}{\textsc{cmms}}

\newcommand{\efg}{\textsc{efg}}

\begin{document} 
	
	\title{Determination of the nature of fluctuations using \textsuperscript{8}{Li} and \textsuperscript{9}{Li} $\beta$-\textsc{nmr} and spin-lattice relaxation.}

	\author{A. Chatzichristos}
			\email{aris.chatzichristos@alumni.ubc.ca}
	\affiliation{Department of Physics and Astronomy, University of British Columbia, Vancouver, BC V6T~1Z1, Canada}
	\affiliation{Stewart Blusson Quantum Matter Institute, University of British Columbia, Vancouver, BC V6T~1Z4, Canada}
	\author{R.M.L. McFadden}
	\affiliation{Stewart Blusson Quantum Matter Institute, University of British Columbia, Vancouver, BC V6T~1Z4, Canada}
	\affiliation{Department of Chemistry, University of British Columbia, Vancouver, BC V6T~1Z4, Canada}
	\author{V.L. Karner}
	\affiliation{Stewart Blusson Quantum Matter Institute, University of British Columbia, Vancouver, BC V6T~1Z4, Canada}
	\affiliation{Department of Chemistry, University of British Columbia, Vancouver, BC V6T~1Z4, Canada}
	\author{D.L. Cortie}
	\affiliation{Department of Physics and Astronomy, University of British Columbia, Vancouver, BC V6T~1Z1, Canada}
	\affiliation{Stewart Blusson Quantum Matter Institute, University of British Columbia, Vancouver, BC V6T~1Z4, Canada}
	\affiliation{Department of Chemistry, University of British Columbia, Vancouver, BC V6T~1Z4, Canada}
	\author{C.D.P. Levy}
	\affiliation{TRIUMF, 4004 Wesbrook Mall, Vancouver, BC V6T~2A3, Canada}
	\author{W.A. MacFarlane}
	\affiliation{Stewart Blusson Quantum Matter Institute, University of British Columbia, Vancouver, BC V6T~1Z4, Canada}
	\affiliation{Department of Chemistry, University of British Columbia, Vancouver, BC V6T~1Z4, Canada}
	\author{G.D. Morris}
	\affiliation{TRIUMF, 4004 Wesbrook Mall, Vancouver, BC V6T~2A3, Canada}
	\author{M.R. Pearson}
	\affiliation{TRIUMF, 4004 Wesbrook Mall, Vancouver, BC V6T~2A3, Canada}
	\author{Z. Salman}
	\affiliation{Laboratory for Muon Spin Spectroscopy, Paul Scherrer Institute, CH-5232 Villigen PSI, Switzerland}
	\author{R.F. Kiefl}
	\affiliation{Department of Physics and Astronomy, University of British Columbia, Vancouver, BC V6T~1Z1, Canada}
	\affiliation{Stewart Blusson Quantum Matter Institute, University of British Columbia, Vancouver, BC V6T~1Z4, Canada}
	\affiliation{TRIUMF, 4004 Wesbrook Mall, Vancouver, BC V6T~2A3, Canada}
	
	\date{\today}

	\begin{abstract}
	We report a comparison of the $1/T_1$ spin lattice relaxation rates (\textsc{slr}) for \textsuperscript{9}{Li} and \textsuperscript{8}{Li} in Pt and SrTiO$_{3}$, in order to differentiate between magnetic and electric quadrupolar relaxation mechanisms. In Pt, the ratio of the $1/T_{1}$ spin relaxation rates $R_{Pt}$ was found to be 6.82(29), which is close to but less than the theoretical limit of $\sim7.68$ for pure magnetic relaxation. In SrTiO$_{3}$ this ratio was found to be 2.7(3), which is close but larger than the theoretical limit of $\sim2.14$ expected for pure electric quadrupolar relaxation. These results bring new insight into the nature of the fluctuations in the local environment of implanted \textsuperscript{8}{Li} observed by $\beta$-\textsc{nmr}.
	\end{abstract}
	
	\maketitle

	\section{Introduction \label{sec:introduction}}
	
	\eli\ $\beta$-detected \nmr\ (\bnmr) has been established as a powerful tool for material science due to its inherent sensitivity to magnetic and electronic properties~\cite{2015-MacFarlane-SSNMR-68-1}. The principal success of \triumf's low-energy incarnation of \bnmr~\cite{2014-Levy-HI-225-165,2014-Morris-HI-225-173} is the ability to study thin films, surfaces, and interfaces --- where conventional \nmr\ is difficult or impossible. This stems from \bnmr's high  sensitivity  relative to conventional \nmr; for \bnmr\ typically only $\sim10^{8}$ nuclei (instead of $\sim 10^{17}$) are required for a  signal. The only other real-space technique with equivalent sensitivity over a comparable material length scale (viz. \SIrange{10}{200}{\nano\metre})~\cite{Lee20151} is low-energy \musr\ (\lemusr)~\cite{2004-Bakule-CP-45-203} ; however, it operates in a complementary time-window due to the different probe lifetimes (\SI{1.21}{\second} for \elip\ vs. \SI{2.2}{\micro\second} for $\mu^{+}$). Thus, both techniques have leveraged the nuclear physics of beta decay to investigate topical problems in condensed matter physics including magnetic surfaces, thin film heterostructures, topological insulators, superconductors etc.
	
	A key issue in any \eli\ \bnmr\ experiment is to identify the source of spin-lattice relaxation (\slr) and in particular whether the fluctuations driving the \slr\ are magnetic or electric in origin. Unlike the positive muon, $\mu^{+}$ ($I=1/2$), \eli\ ($I=2$) is \emph{not} a pure magnetic probe and its relaxation is sensitive to both fluctuating magnetic fields and electric field gradients (\efg's). In some cases, the primary source of relaxation may be inferred. For example, in simple metals the observed relaxation is linear in temperature~\cite{2009-Hossain-PB-404-914} as expected from the Korringa relaxation~\cite{1951-Korringa-P-7-601}, which originates from a magnetic hyperfine interaction between the nuclear spin and the spin of the conduction electrons.  However, in more complicated instances, such as heterostructures comprised of magnetic and non-magnetic layers, it becomes difficult to determine the contribution of each type of relaxation. \lao/\sto\ multilayers are particularly illustrative of this point; the bulk layers are non-magnetic insulators, while there is evidence of magnetism at their interfaces~\cite{2012-Salman-PRL-109-257207}.
	
In conventional \nmr\ it is possible to differentiate between relaxation mechanisms by isotopic variation of the nuclear probe, since the absolute relaxation rates for each isotope scale according to their nuclear moments. For two isotopes with significantly different nuclear moments (e.g., \textsuperscript{6}Li and \textsuperscript{7}Li~\cite{1998-Tomeno-JPSJ-67-318}) the ratio of the relaxation rates should be distinctly different in the limits of either pure magnetic or pure electric quadrupolar relaxation. In this study we test the feasibility of isotope comparison applied to \bnmr\ --- using \eli\ and \nli, two $\beta$-radioactive isotopes. The stopping sites of \eli\ and \nli\ are often interstitial rather than substitutional as in the case of conventional  NMR. However, we expect that both implanted \eli\ and \nli\ will probe the same sites. Measurements on \nli\ are more time consuming than those for \eli. This is related to the fact that \nli\ lies one neutron further away from the valley of stability, consequently the beam intensity in this experiment was about 10 times lower for \nli\ than \eli, and has a more complicated $\beta$-decay scheme, which results into a beta-decay asymmetry for \nli\ about 3 times smaller than for \eli, as will be discussed below. 
	
Measurements reported here were made in Pt metal, where the spin relaxation rate of \eli\ (\nli) is dominated by Korringa scattering~\cite{2012-Ofer-PRB-86-064419}, which is magnetic, and in strontium titanate (\sto), which is a non-magnetic insulator with a large static electric quadrupolar interaction for implanted \eli. \sto\ is a common substrate material but also has interesting properties on its own which have been  studied extensively with a wide variety of methods including \bnmr. Although we expect the quadrupolar fluctuations in \efg\ causing spin relaxation to dominate, there are also potential magnetic sources of relaxation that could contribute as explained below.
		  
	In the following sections we first summarize the theoretical considerations behind \bnmr, as well as the isotopic variation method. This is then followed by a description of the experiment and finally we present the experimental results along with a discussion.     
	  
	\section{Theory \label{sec:theory}}
	
	The basis of \bnmr\ is the parity-violating weak interaction, whereby the direction of the emitted electron (positron) from the decaying nucleus is correlated with the nuclear spin polarization at the time of decay:
		
	\begin{equation}\label{eq:angle}
	W(\theta) = 1 + \beta ap \cos(\theta)
	\end{equation}
	where $\beta=\nu/c$ is the velocity of the high energy electron (positron)  normalized to the speed of light, $p$ is the magnitude of the nuclear polarization vector, $\theta$ is the angle between the nuclear polarization and the electron (positron) velocity   and $a$ is the asymmetry parameter depending on the properties of nuclear $\beta$-decay. The theory of nuclear beta-decay predicts that $a$ is about 1/3 for \eli\ and considerably smaller ($\sim0.1$) for \nli~\cite{Firestone1996}, if averaged over all the decay modes.
		
\begin{figure}
				\centering
				\includegraphics[width=0.45\textwidth]{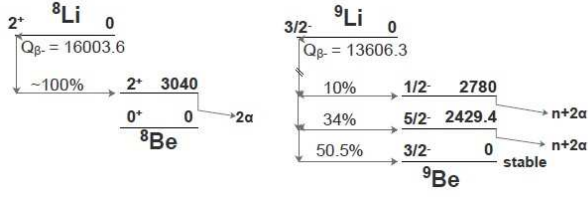}
				\caption{\label{fig:decay} Properties of the principle $\beta$-decay modes of \eli\ and \nli~\cite{Firestone1996}. The asymmetry of each decay mode of \nli\ is documented in the table below. The total asymmetry for \nli\ is the sum of the asymmetry weighted by the relevant probability of each decay mode.}
				\begin{ruledtabular}
					\begin{tabular}{r l l l l l}
						$^{9}$Be state & Probability & $I^{\pi}$ & $a$ &  Decay mode  \\
						\hline
						ground state	& 50.5$\%$ & $3/2^{-}$ & -2/5 & stable \\
						2429.4 MeV & 34$\%$ & $5/2^{-}$	 & 3/5  & n+2$\alpha$	\\
						2780 MeV	& 10$\%$ & $1/2^{-}$ & -1 & n+2$\alpha$ \\
					\end{tabular}
				\end{ruledtabular}
			\end{figure}

	The reduction in asymmetry for \nli\ compared to \eli\ is attributed to \nli's more complicated $\beta$-decay scheme. In particular, \nli\ has three main decay channels, two of which have opposite asymmetries that nearly cancel after weighting by the branching probabilities. Thus, most of the observed asymmetry is from the weakest decay mode which has a branching probability of only \num{0.1} but a large theoretical asymmetry parameter $a=1.0$. The relevant branching probabilities  and asymmetries of each decay mode are reported in Fig.~\ref{fig:decay}. We note in passing that it should be possible to enhance the \bnmr\ signal from \nli\ by tagging events according to whether an $\alpha$ is emitted or not, which will allow us to distinguish between the different decay channels and isolate their contributions. This is currently being explored as a way to optimize the \bnmr\ of \nli. 
	
The resulting  anisotropic decay pattern for the high energy electron (positron) allows one to monitor the nuclear polarization from highly polarized \elip\ or \nlip\ beams implanted in the sample. In particular, the asymmetry in the count rate at time $t$  between two opposing beta detectors is proportional to the component of nuclear polarization along the direction defined by the two detectors: 
		
	\begin{equation} \label{eq:asymmetry}
	A(t)= \frac{N_{B}(t) -N_{F}(t) }{N_{B}(t)+N_{F}(t) }=A_0p_z(t)
	\end{equation}
	where $N_{B}(t)$ and $N_{F}(t)$ are the counts measured in the backward and forward detectors, $p_z(t)$ is the component of nuclear polarization along the $z$-axis defined by the detectors, and $t$ is the time of decay after implantation. The detectors are generally positioned so that $z$ is along the direction of initial polarization. Note that the asymmetry in the count rate has a maximum value of $A_0$ at $t=0$ which is reduced relative to the theoretical asymmetry $a$, as calculated from the nuclear properties, owing to instrumental effects such as the finite solid angle subtended by the detectors and scattering of the betas before reaching the detectors. Note also that $p_z(t)$ and thus $A(t)$, are time dependent, reflecting the fact that the nuclear polarization is subject to spin relaxation processes in the sample, which in fact is the quantity of interest in this experiment.    

Information on the fluctuations of the electromagnetic fields in a material of interest is obtained through measurements of the spin-lattice relaxation (\slr) rate in the absence of a \rf\ magnetic field. The \slr\ may be studied by implanting a series of beam pulses into the sample and then monitoring $\mathcal{A}(t)$, which is the convolution of $A(t-t')$  with the beam pulse $N(t')$ where $t'$ is the time of arrival for given probe and $t-t'$ is the time spent in the sample before its beta decay:

\begin{equation} \label{eq:calA}
	\mathcal{A}(t)= \int_{-\infty}^t N(t')A(t-t')dt'
	\end{equation}	
	
	In general the \slr\ rate, usually denoted as $1/T_{1}$ (with $T_{1}$ being the longitudinal spin-lattice relaxation time), originates from fluctuations in the local environment arising from  fundamental processes such as phonon scattering, magnon scattering, conduction electron scattering, diffusion, etc. The total observed rate can be decomposed into a sum of individual contributions, which may be grouped into magnetic ($1/T_{1}^{\mathrm{M}}$) and electric quadrupolar ($1/T_{1}^{\mathrm{Q}}$) terms:
	\begin{equation} \label{eq:rate}
	\frac{1}{T_{1}} = \frac{1}{T_{1}^{\mathrm{M}}} + \frac{1}{T_{1}^{\mathrm{Q}}},
	\end{equation}
	 Most often one of the relaxation mechanisms will dominate. For instance, we expect Korringa relaxation to be dominant in simple metals. 
	
The magnitudes of each contribution for a given probe nucleus scale according to their nuclear properties; namely, their spin, $I$, magnetic moment, $\mu$, and electric quadrupole moment, $Q$. Measurements of \slr\ rates for two different isotopes under identical experimental conditions (i.e., magnetic field, temperature, etc.) can be compared through their ratio, $R$:
	\begin{equation} \label{eq:ratio}
	R\left(I,I^{\prime}\right) \equiv \frac{1/T_{1}(I)}{1/T_{1}(I^{\prime})} = \frac{ 1/T_{1}^{\mathrm{M}}(I) + 1/T_{1}^{\mathrm{Q}}(I) }{ 1/T_{1}^{\mathrm{M}}(I^{\prime}) + 1/T_{1}^{\mathrm{Q}}(I^{\prime}) },
	\end{equation}
	where $I$ and $I^{\prime}$ denote the spin quantum number of each isotope. Two limits are of interest here: when the relaxation is \emph{solely} due to either magnetic or quadrupolar interactions within the host-sample.
	In the former case, Eq.~\eqref{eq:ratio} reduces to the ratio of pure magnetic relaxation, $R_{\mathrm{M}}$, which in the limit of fast fluctuations (i.e., $\tau^{-1}_{c}\gg\omega_{0}$, where $\tau_{c}$ is the \nmr\ correlation time and $\omega_{0}$ is the Larmor resonance frequency) is:  
	\begin{equation} \label{eq:ratio-magnetic}
	R_{\mathrm{M}}\left(I,I^{\prime}\right) = \left( \frac{\mu/I}{\mu^{\prime}/I^{\prime}} \right)^{2} = \left ( \frac{\gamma}{\gamma^{\hspace{2 pt}\prime}} \right )^{2},
	\end{equation}
	where $\mu$ and $\gamma$ are the magnetic moment and gyromagnetic ratio of each isotope. Note that the fast fluctuation limit ensures that $1/T_{1}$ is independent of $\omega_{0}$. 
	
	In the other case, Eq.~\eqref{eq:ratio} yields the ratio of relaxation rates in the pure quadrupolar limit, $R_{\mathrm{Q}}$:
	\begin{equation} \label{eq:ratio-quadrupolar}
	R_{\mathrm{Q}}\left(I,I^{\prime}\right) = \frac{f(I)}{f(I^{\prime})} \left ( \frac{Q}{Q^{\prime}} \right )^{2},
	\end{equation}
	where $Q$ are the nuclear quadrupole moments, and~\cite{1961-Abragam-PNM}
	\begin{equation} \label{eq:spin-factor}
	f(I) = \frac{2I+3}{I^{2}(2I-1)} 
	\end{equation}
	
	Thus, given  the nuclear moments of each isotope, one can calculate the ratio of relaxation rates when either mechanism is dominant. Using Eqs.~\eqref{eq:ratio-magnetic} and \eqref{eq:ratio-quadrupolar}, along with the nuclear spins and moments for \eli\ and \nli\ (see Table~\ref{tab:isotopes}), we find the limiting cases for $T_{1}^{-1}(^{9}\mathrm{Li})/T_{1}^{-1}(^{8}\mathrm{Li})$: \num{7.67964 \pm 0.00016} and \num{2.1362 \pm 0.0004} for $R_{\mathrm{M}}$ and $R_{\mathrm{Q}}$, respectively. The difference between these limits is not as pronounced as for \textsuperscript{6}Li and \textsuperscript{7}Li~\cite{1998-Tomeno-JPSJ-67-318}, where $R_{\mathrm{M}}$ and $R_{\mathrm{Q}}$ differ by a factor of \num{\sim 90}~\cite{2005-Stone-ADNDT-90-75}. Nevertheless, \eli\ and \nli\ are sufficiently different that the  nature  of fluctuations and resulting spin relaxation (magnetic versus electric quadrupolar) may be differentiated by such a comparison.
	
	\begin{table*}
	\caption{\label{tab:isotopes}Intrinsic nuclear properties of Li radioisotopes used in \bnmr\ and \bnqr. $I^{\pi}$ is the nuclear spin (and parity), $\mu$ is the magnetic moment, and $Q$ is the electric quadrupole moment.}
		\begin{ruledtabular}
			\begin{tabular}{r l l l l l}
				& $I^{\pi}$	& $\tau_{\beta}$ (s) &  $\mu$ ($\mu_{\mathrm{N}}$)\footnotemark[1]  & $Q$ (mb)\footnotemark[2]	& \\
				\hline
				\eli	& $2^{+}$	& 1.2096(5)~\cite{2010-Flechard-PRC-82-027309} & +1.653560(18)~\cite{2006-Borremans-PRC-72-044309} & +32.6(5)~\cite{2011-Voss-JPGNPP-38-075102} \\
				\nli 	& $3/2^{-}$	& 0.2572(6)~\cite{1976-Alburger-PRC-13-835}    & +3.43678(6)~\cite{2006-Borremans-PRC-72-044309} & -31.5(5)~\cite{2011-Voss-JPGNPP-38-075102} \\
			\end{tabular}
			\footnotetext[1]{The magnetic moments have been corrected for diamagnetic shielding.}
			\footnotetext[2]{The quadrupole moments were determined from their ratios, starting with the well-known value for \textsuperscript{7}Li~\cite{2005-Stone-ADNDT-90-75}.}
		\end{ruledtabular}
	\end{table*}
	
	\section{Experimental \label{sec:experimental}}
	
	The experiment was performed using \SI{18}{\kilo\electronvolt} beams of \elip\ and \nlip\ at \triumf's Isotope Separator and Accelerator Facility (ISAC) in Vancouver, Canada. \isac\ is capable of providing an intense beam for a large number of isotopes of various elements~\cite{isac-web}, including \eli\ and \nli. For this experiment, \triumf's dedicated \bnmr\ and \bnqr\ spectrometers were used. A detailed discussion on the characteristics of the spectrometers can be found elsewhere~\cite{2014-Morris-HI-225-173, 2004-Morris-PRL-93-157601}.
	
	Before reaching the spectrometer, the Li$^{+}$ ion beam first passes through the ISAC polarizer~\cite{2014-Levy-HI-225-165}. The first stage of the polarizer is to neutralize the beam by passing it through a Rb vapor cell. The neutral beam then drifts  $\sim2$m during which time the $^2S_{1/2}-^2P_{1/2}$ optical $D_{1}$ transition is pumped with circularly polarized laser light. The last stage is to re-ionize the beam in a He gas so that the polarized beam can be delivered alternately to the spectrometers. Previous work shows that the nuclear polarization of the  beam after stopping in the sample is $\sim 70\%$ \cite{2014-MacFarlane-JPCS-551-012059}. 
	
	It is important to note that unlike conventional \nmr, where the Boltzman factor determines the polarization, the nuclear polarization in \bnmr\ is close to unity and independent of the sample temperature and magnetic field. Consequently, measurements can be made under conditions where conventional \nmr\ is difficult or impossible e.g. at high temperatures, low magnetic fields or in thin films. The intensity of the implanted beam (typically \SI{\sim e7}{\per\second}), is such that the concentration of the nuclear probes is so small that there is no interaction  between probes and thus no homonuclear spin-coupling. 
	
	\section{Results and Discussion \label{sec:results}}
	
	To demonstrate the comparison of \eli\ and \nli\ in \bnmr, two very different materials were selected. The first is Pt which is a d-band metal in which the \eli\ resides at a site with little or no quadrupolar interaction. In this test case we expect the relaxation to be predominantly magnetic, originating from Korringa scattering. \sto\ on the other hand is a non-magnetic insulator with few nuclear moments. Previous work in \sto\ shows that \eli\ experiences a large quasistatic quadrupolar interaction~\cite{2003-MacFarlane-PB-326-209}. Thus in this case we expect quadrupole fluctuations to play a more important role. Nevertheless, it is still unclear to what extent magnetic effects can be neglected in \sto. For example, it is well known that the implantation of \eli\ generates vacancy-interstitial pairs as well as electron-hole pairs. Such defects are often magnetic. For example O vacancies in \sto\ result in two Ti$^{3+}$ ions which are typically paramagnetic. In principle, the resulting paramagnetic defects would have low frequency magnetic fluctuations which will contribute to the \slr\ of the implanted Li nucleus in \sto. 
	
	\subsection{Platinum \label{sec:pt}}
	
	\elip\ resonance measurements in Pt have shown a single narrow line below \SI{300}{\kelvin}, indicating that \elip\ occupies a single site with a vanishing (static) \efg~\cite{2009-Fan-PB-404-906,2012-Ofer-PP-20-156}. The spectrum is also simpler than in other metals, where multiple \lip\ sites are found below \SI{300}{\kelvin}~\cite{2003-MacFarlane-PB-326-213,2007-Parolin-PRL-98-047601,2007-Salman-PRB-75-073405,2008-Parolin-PRB-77-214107,2009-Wang-PB-404-920,2009-Parolin-PRB-80-174109,2012-Chow-PRB-85-092103}.
	
\begin{figure*}
		\centering
		\includegraphics[width=0.9\textwidth]{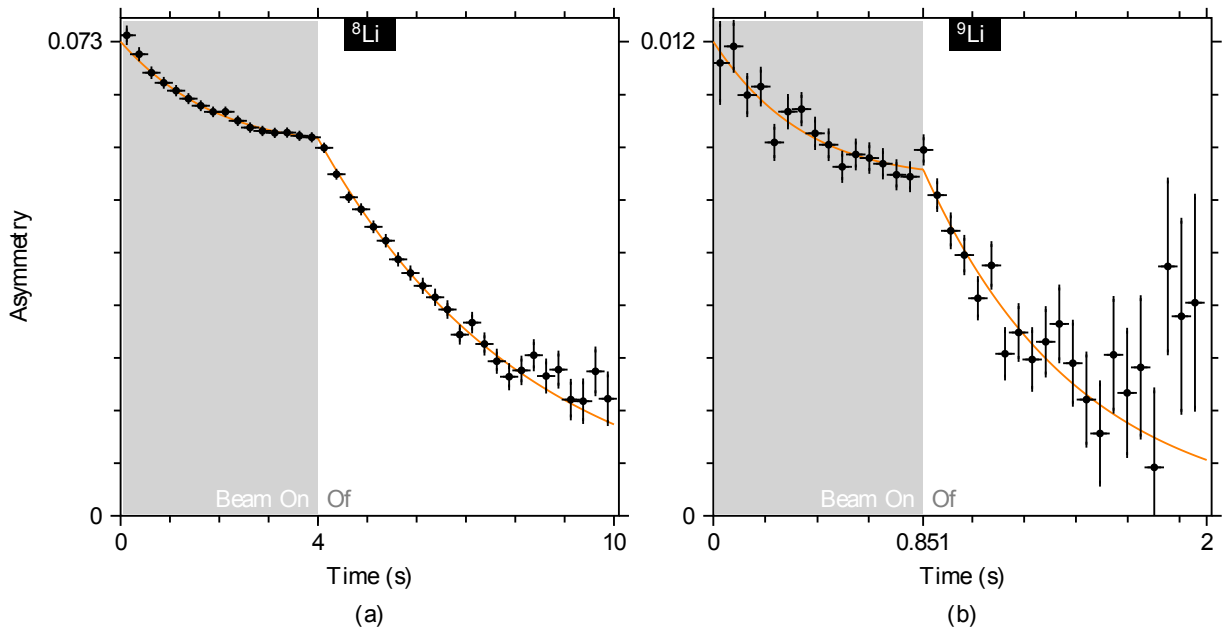}
		\caption{\label{fig:8li-9li-slr-65kG} \slr\ spectra for \elip\ (left) and \nlip\ (right) implanted in Pt foil with an energy of \SI{18}{\kilo\electronvolt} at \SI{300}{\kelvin} under \SI{6.55}{\tesla}. The solid lines are fits to Eq.~\eqref{eq:slr2}. Note the different time scales, which reflect the lifetime of each radionuclide. The absolute \slr\ rate for \nlip\ is \SI{1.60 \pm 0.1} and \SI{0.2368 \pm 0.0026} for \elip.}
	\end{figure*}

Given the simplicity of the resonance spectrum we expect \slr\ in Pt to follow a single exponential form with:
\begin{equation} \label{eq:exponential}
A(t-t^{\prime}) = \exp \left [ -\lambda \left ( t - t^{\prime} \right)/T_1 \right ],
\end{equation}
 Substituting this into Eq.~\eqref{eq:calA} and assuming a square beam pulse during the time interval [0,$\Delta$], one obtains a form for the asymmetry during and after the pulse given by:  
\begin{equation} \label{eq:slr2}
\mathcal{A}(t)=\begin{cases} A_0\frac{\tau'}{\tau_{\beta}}\frac{1-\exp(-t/\tau')}{1-\exp(-t/\tau_{\beta})}& t\leq \Delta \\
A(\Delta)\exp[-(-t-\Delta)/T_1] & t> \Delta
\end{cases},
\end{equation}
where $\tau_{\beta}$ is the radioactive lifetime, $1/\tau^{\prime}=1/\tau_{\beta}+ 1/T_{1}$ and $A_{0}$ is the initial asymmetry at  the time of implantation. Note that the \slr\ spectrum has two distinct regions (see Fig.~\ref{fig:8li-9li-slr-65kG}): during the beam pulse ($0<t<\Delta$) the asymmetry relaxes towards a dynamic equilibrium value \cite{2009-Hossain-PB-404-914}: 
\begin{equation} \label{eq:equillibrium}
	\bar{\mathcal{A}} = \frac{A_{0}}{1+\tau_{\beta}/T_{1}},
	\end{equation}	
whereas after the beam pulse ($t>\Delta$) ${\mathcal A}(t)$ decays towards the Boltzman equilibrium value, which is essentially zero on our scale. Note the pronounced kink in ${\mathcal{A}}(t)$ at $t=\Delta$ when the beam  pulse ends. This is also the time with the highest event rate and smallest statistical uncertainty in ${\mathcal{A}}(t)$. For both isotopes the length of the beam pulse ($\sim 3.3\tau_{\beta}$) and the total observation time (${\sim 9.9}\tau_{\beta}$) were chosen to minimize the statistical uncertainties. 	
	
	The \slr\ rates for \elip\ and \nlip\ implanted at an energy of \SI{18}{\kilo\electronvolt} at \SI{300}{\kelvin} were measured in magnetic fields of \SI{1.90}{\tesla} and \SI{6.55}{\tesla} --- the latter shown in Fig.~\ref{fig:8li-9li-slr-65kG}. Several general distinctions should be pointed out between \slr\ spectra for \elip\ and \nlip\ in Pt: The initial asymmetry (i.e., $A_{0}$) for \elip\ is \num{\sim 6} times greater than for \nlip; $1/T_{1}$ is \num{\sim 7} times larger for \nlip\ than for \elip; and the relative uncertainty of the \slr\ rate measurements for \nlip\ is greater by a factor of \num{\sim 5} than for \elip. The latter can be understood as follows: The statistical figure of merit for any \bnmr\ measurement is $A^{2}N$, where $A$ is the observable asymmetry and $N$ is the total number of decay events --- both factors for \nli\ are significantly reduced relative to \eli. Since \nli\ lies further away from the valley of nuclear stability, it has a shorter half-life and fewer ions are extracted from the ion source and delivered to the spectrometer (here \SI{\sim e6}{\per\second} vs \SI{\sim e7}{\per\second} for \elip). This in turn reduces the factor $N$ for \nli. Also, as explained above, the asymmetry for \nli\ is much smaller than for \eli. As a result, about 90\% of the data acquisition  was spent on \nli, since these results dominated our uncertainty in the ratio of the relaxation rates. 
	
	\begin{figure}
		\centering
		\includegraphics[width=0.45\textwidth]{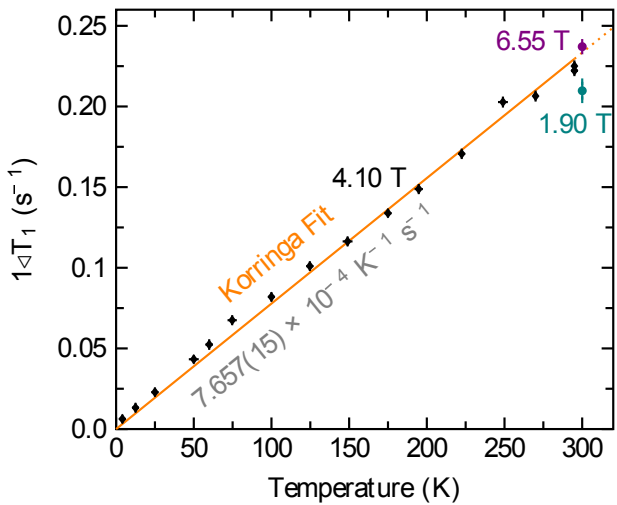}
		\caption{\label{fig:pt-slr-korringa} Measured \slr\ rates for \eli\ implanted in Pt. The relaxation rate increases linearly with temperature, appearing insensitive to both implantation energy and magnetic field strength, consistent with a Korringa mechanism~\cite{1951-Korringa-P-7-601}. Measurements from this work are highlighted in colored disks, while black diamond markers indicate data from earlier measurements on Pt foil~\cite{2012-Ofer-PRB-86-064419}. The solid line is Korringa fit to \emph{all} the \slr\ rates in Pt and differs somewhat from the result of \citeauthor{2012-Ofer-PRB-86-064419} due to the additional data points from this work.}
	\end{figure}
	
	Temperature dependent \slr\ of \elip\ in Pt has been studied previously by \citeauthor{2012-Ofer-PRB-86-064419} between \SIrange{3}{295}{\kelvin} at \SI{4.10}{\tesla}, where the \slr\ rate was found to increase linearly with temperature, implying Korringa relaxation~\cite{1951-Korringa-P-7-601}. This relation holds for high magnetic fields and different implantation energies. The temperature-dependent \elip\ \slr\ rates at various magnetic fields are shown in Fig.~\ref{fig:pt-slr-korringa}, including our measurements, as well as results on Pt foil by \citeauthor{2012-Ofer-PRB-86-064419}. The \eli\ \slr\ rate at \SI{6.55}{\tesla} is in good agreement with the Korringa fit by \citeauthor{2012-Ofer-PRB-86-064419}, extrapolated to \SI{300}{\kelvin}, whereas the measured \slr\ rate at \SI{1.9}{\tesla} is lower by about 10\%. It is unlikely that this is a real effect since any additional source of relaxation  would {\it increase} the relaxation at the lower magnetic field which is opposite to what is observed. The slight reduction in $1/T_1$ measured at 1.9 T suggests there may be  a small systematic error related to the fact that the beam spot is a bit larger and the ratio between the beta rates in the two detectors is different compared to the higher field. However, it should be noted that the measured \eli\ \slr\ rates in Pt foil appear to increase linearly with temperature, independent of implantation energy and applied magnetic field. 

The ratios of $T_{1}^{-1}(^{9}\mathrm{Li})/T_{1}^{-1}(^{8}\mathrm{Li})$ at \SI{6.55}{\tesla} and \SI{1.90}{\tesla} are in good agreement with each other and we find a relaxation rate ratio, $R_{\mathrm{Pt}}$, of \SI{6.8 \pm 0.4} and \SI{5.9 \pm 0.9} at \SI{6.55}{\tesla} and \SI{1.90}{\tesla}, respectively.

	\subsection{Strontium Titanate \label{sec:sto}}
	
	\sto\ was chosen as a second example, since it is a non-magnetic insulator and  representative of a material where the relaxation in \eli\ \bnmr\ is expected to be dominated by quadrupolar fluctuations. It has been studied extensively with low-energy \eli\ \bnmr~\cite{2009-Smadella-PB-404-924,2011-Salman-PRB-83-224112}. Implanted \eli\ occupies a single non-cubic site, which is unambiguously evidenced by the observation of a pure nuclear quadrupole resonance (\nqr) in zero magnetic field~\cite{2004-Salman-PRB-70-104404,2006-Salman-PB-374-468,2006-Salman-PRL-96-147601,2014-Voss-JPGNPP-41-015104}.
	
	\begin{figure*}
		\centering
		\includegraphics[width=0.9\textwidth]{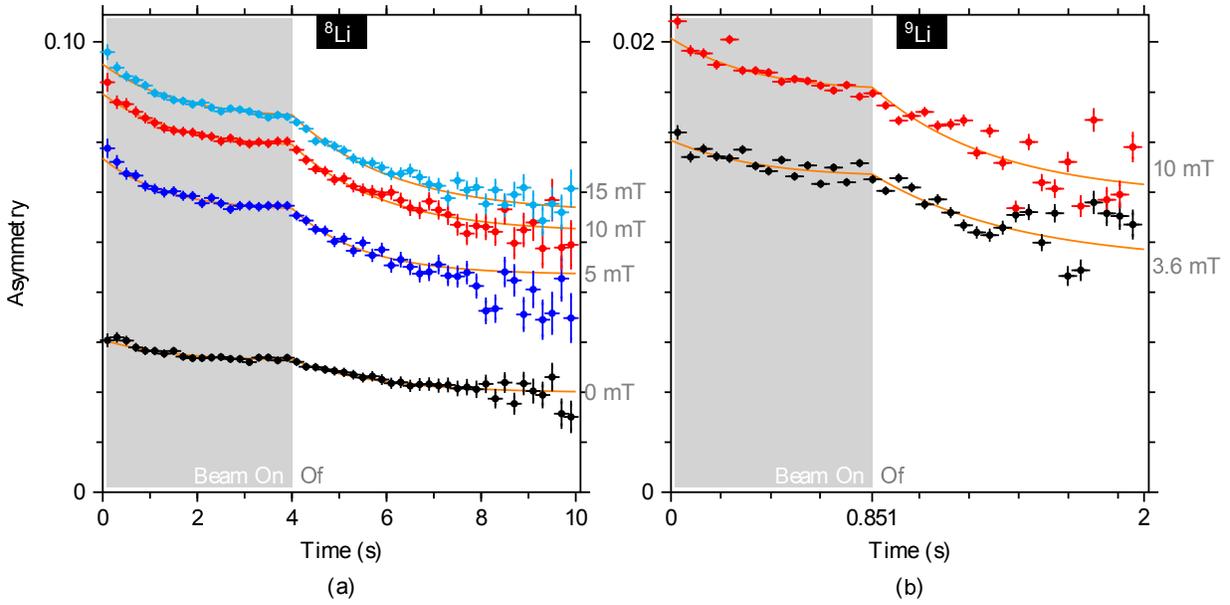}
		\caption{\label{fig:sto-slr-spectra} \slr\ spectra of \eli\ (\emph{left}) and \nli\ (\emph{right}) in single crystal \sto\ at \SI{300}{\kelvin}. The solid lines are a global fit to Eqs. \eqref{eq:calA} and \eqref{eq:exponential-nonrelax} where a common parameter $f$ is shared between all spectra.}
	\end{figure*}
	
	Figure \ref{fig:sto-slr-spectra} shows the \slr\ spectra for \eli\ and \nli\ at \SI{300}{\kelvin} at various magnetic fields between \SIrange{0}{15}{\milli\tesla} applied along a (1 0 0) cubic crystallographic axis. It is evident from the data that the relaxation is more complex than in Pt since a single exponential fails to  describe the decay of spin polarization. One might expect this since a magnetic field applied along the (1 0 0) direction breaks the local symmetry between the 3 otherwise equivalent sites. More specifically the \efg\ tensor is axially symmetric about one of the three orthogonal cubic axes. Thus the applied magnetic field is either along \efg\ axis or perpendicular to it. However the relaxing fraction, $f$, was approximately field-independent, within our range of fields including the spectrum at zero field. Given the two 90 degree sites don't contribute to the ZF signal and that $f$ is about the same at ZF, the more complex relaxation function observed in \sto\ must be unrelated to the angle between the magnetic field and the symmetry axis of the \efg. Consequently there must be an additional source of fluctuations affecting the \slr\ for all 3 sites in the same way but in an inhomogeneous manner either in time or space. This could be due to the dynamics associated with defects close to some of the implanted Li. In any event,  given one of the relaxation rates is found to be nearly zero, a phenomenological relaxation function of the following form~\cite{2009-Smadella-PB-404-924} was used:
	\begin{equation} \label{eq:exponential-nonrelax}
	A(t-t^{\prime}) = f\exp \left [ -\lambda \left ( t - t^{\prime} \right) \right ] + \left ( 1 - f \right ),
	\end{equation}
	where $f$ is the fraction of the relaxing asymmetry ($0 \leq f \leq 1$) and $\lambda \equiv 1/T_{1}$. 
	
	\begin{figure}
		\centering
		\includegraphics[width=0.45\textwidth]{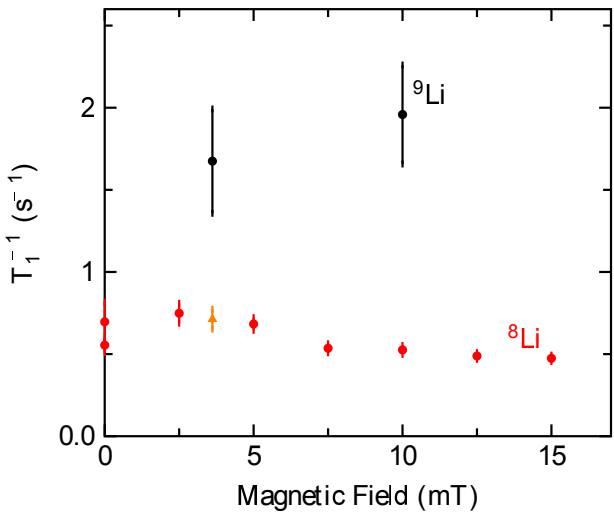}
		\caption{\label{fig:sto-slr-field}Field dependence of $1/T_{1}$ for \eli\ and \nli\ in \sto\ at \SI{300}{\kelvin}. The (orange) triangle represents a linear interpolation at \SI{3.6}{\milli\tesla} from the \SI{2.5}{\milli\tesla} and \SI{5}{\milli\tesla} \eli\ measurements.}
	\end{figure}
	
	Regarding the relaxation function, note that this is an unfamiliar regime, where the Zeeman interaction is smaller than $\nu_{Q}=\SI{153.2}{\kilo\Hz}$ over the full range of fields, since even for our highest field measurement at \SI{15}{\milli\tesla}, $\gamma B=\SI{94}{\kilo\Hz}$. At high fields (several Tesla), previous work suggests that $f=0$~\cite{2003-MacFarlane-PB-326-209}, so there is likely some change that will happen around the area of \SI{50}{\milli\tesla}, where the Zeeman interaction really starts to take over.
	
Since $f$ is approximately field-independent, the \slr\ spectra for \eli\ and \nli\ were fit globally, sharing a common $f$, which turned out to be \num{0.347 \pm 0.003}. The relaxation rates are plotted as a function of applied field in Fig.~\ref{fig:sto-slr-field}.   
	
	The \slr\ rate for \eli\ is found to vary weakly with applied magnetic field below \SI{15}{\milli\tesla}, reaching a plateau below \SI{5}{\milli\tesla} (see Fig.~\ref{fig:sto-slr-field}). It is likely, but unclear due to the limited statistics, that a similar behavior occurs for \nli. At \SI{300}{\kelvin}, the ratio of the \nli/\eli\ \slr\ rates for \sto, $R_{\mathrm{STO}}$, was found to be \num{3.7 \pm 0.7} at \SI{10}{\milli\tesla} and \num{2.4 \pm 0.5} at \SI{3.6}{\milli\tesla}.
	
	For comparison, the \slr\ rate of \eli\ and \nli\ was also measured in a second \sto\ sample \footnote{This sample was ${\mathrm{Sr}}{\mathrm{TiO}}_{3}$ with \SI{30}{\nano\meter} ${\mathrm{La}}{\mathrm{TiO}}_{3}$ capping layer. Bulk ${\mathrm{La}}{\mathrm{TiO}}_{3}$ is a prototypical Mott insulator and is antiferromagnetic below \SI{\sim 135}{\kelvin}. Like ${\mathrm{Sr}}{\mathrm{TiO}}_{3}$, \textbf{${\mathrm{La}}{\mathrm{TiO}}_{3}$} is non-magnetic at \SI{300}{\kelvin}, though transport measurements have shown the existence of a metallic and superconducting heterointerface; however, at an implantation energy of \SI{18}{\kilo\electronvolt}, only a negligible fraction of the probing ions stops in the ${\mathrm{La}}{\mathrm{TiO}}_{3}$ film, or near the interface. This was confirmed for both $\textsuperscript{8}{Li}$ and $\textsuperscript{9}{Li}$ by using the Monte Carlo-based simulation package SRIM 2008. For this reason, this can be considered to be effectively an SLR measurement in the ${\mathrm{Sr}}{\mathrm{TiO}}_{3}$ substrate of the film.}. These spectra were fitted globally with the same fitting function as in the first \sto\ sample.
	The shared relaxing fraction in this case was \num{0.341 \pm 0.002}, very close to the value calculated independently from the other \sto\ sample. The ratio of relaxation rates in this sample at \SI{10}{\milli\tesla} was found to be \num{2.4 \pm 0.5}.

	\subsection{Ratio of relaxation rates \label{sec:ratio}}
	
	\begin{figure}
		\centering
		\includegraphics[width=0.45\textwidth]{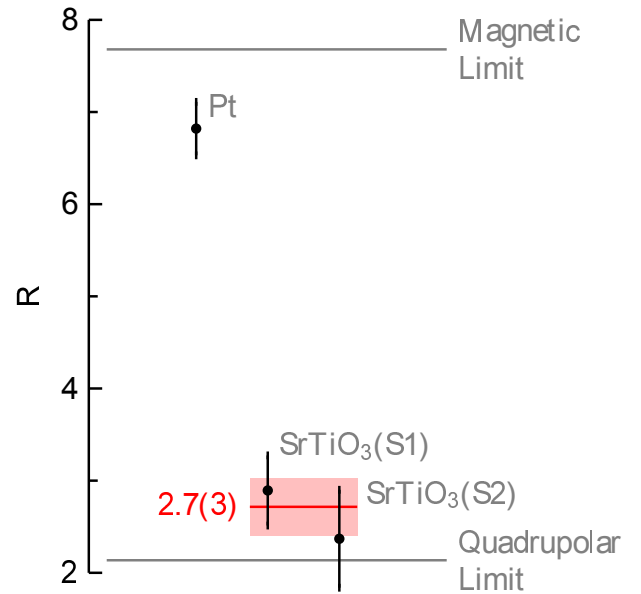}
		\caption{\label{fig:relaxation-ratio} Ratios of \nli\ to \eli\ $1/T_{1}$ relaxation rates in Pt (weighted average of all measurements) and in the two \sto\ samples. The red line represents the weighted average of the measurements in both \sto\ samples.}
	\end{figure}

	The ratio of relaxation rates in platinum $R_{Pt} = \num{6.82 \pm 0.29}$, which is the weighted average of the measurements at \SI{6.55}{\tesla} and \SI{1.90}{\tesla}. Note that this value is somewhat  less than  expected from the pure magnetic limit $R_{M}$ (Fig.~\ref{fig:relaxation-ratio}).

The reason for this discrepancy is puzzling. All measurements were taken at \SI{300}{\kelvin} where the lithium ions could have some quadrupolar contribution due to local vibrations and scattering of phonons which leads to a fluctuating EFG. However $1/T_{1}$ is very linear in temperature, whereas any such contributions would have a stronger temperature dependence. It would be interesting to repeat the measurements at a lower temperature to check if $R_{\mathrm{Pt}}$ is closer to the magnetic limit or not. In principle the scattering of electrons at the Fermi surface, which is responsible for Korringa relaxation (see Fig.~\ref{fig:pt-slr-korringa}), could also produce a fluctuating EFG and a linear temperature dependence $1/T_1$, which is electric quadrupolar in origin. However, we could not find any calculations of this effect. In any case, an electric quadrupolar contribution to $1/T_1$ cannot be very large in Pt at \SI{300}{\kelvin}.    
	
We also reported a value of $R_{STO}$ in two samples of \sto. In the first sample, the weighted average $R_{STO}$ of the measurements at \SI{3.6}{\milli\tesla} and \SI{10}{\milli\tesla} yielded \num{2.9 \pm 0.4}. This value is close, but not within experimental error of the quadrupolar limit of $R_{Q}\approx2.14$. After taking into account the measurement on the second \sto\ sample, which was \num{2.4 \pm 0.5} at \SI{10}{\milli\tesla}, the weighted ratio of relaxation rates in \sto\ is found to be \num{2.7 \pm 0.3}, closer to the quadrupolar limit. Still there is a small disagreement which suggests some small magnetic contribution to $1/T_1$. 
This may be related to the observed non-exponential relaxation function. If it is due to fluctuations which are inhomogeneous in time or space then nearby defects are likely playing some small role. A small portion of these fluctuations could be magnetic in origin. For example any O vacancies a few lattice sites away would give rise to paramagnetic Ti$^{3+}$ ions. Similarly electron-hole pairs in a triplet state would also be magnetic.     
	
	\section{Conclusions \label{sec:conclusions}}
	
	We have measured  the ratio between $1/T_{1}$ of \nli\ and \eli\ in Pt and \sto\ in order to help identify the nature of the fluctuations responsible for the spin relaxation (i.e., if they are magnetic or electric quadrupolar). In Pt, the relaxation is single exponential and the ratio $R_{Pt}$ was found to be very close to but slightly less than the pure magnetic limit. This is consistent with Korringa relaxation being dominant as suggested by the linear temperature dependence in $1/T_1$ reported previously. Nevertheless the small reduction in $R_{Pt}$ relative to the pure magnetic limit means that excitations causing a fluctuating EFG may provide a small contribution to the observed spin relaxation.  Further measurements at lower temperatures would be needed to verify this. 

In \sto\ at \SI{300}{\kelvin} the results confirm that the dominant source of relaxation is electric  quadrupolar. However, the relaxation function is more complicated involving a relaxing part and a non-relaxing part. This suggests there is some inhomogeneous source of fluctuations/spin relaxation, possibly due to nearby defects. The ratio $R_{STO}$ is close to but slightly larger than the pure quadrupolar limit, indicating that there may be some small magnetic contribution. However, the main source of spin relaxation is quadrupolar. This is consistent with expectations given the large quasi-static nuclear quadrupole interaction. 
	
Most importantly, we have demonstrated that the method of isotope comparison can be used in \bnmr\ to distinguish the nature of the fluctuations responsible for $1/T_1$.  This represents an important new tool for \bnmr, since in many systems there is uncertainty in the source of relaxation that cannot be removed simply by varying experimental parameters. 
	
	\begin{acknowledgments}
		We thank \triumf's \cmms\ for their technical support. This work was supported by: NSERC Discovery Grants to R.F.K. and W.A.M.; and \href{http://isosim.ubc.ca/}{IsoSiM} fellowships to A.C. and R.M.L.M. \triumf\ receives federal funding via a contribution agreement with the National Research Council of Canada. 
	\end{acknowledgments}
	
	\bibliography{./references/refs}

\end{document}